\documentclass[aps,prl,twocolumn,showpacs,floats,superscriptaddress,floatfix,longbibliography,hidelinks]{revtex4-1}

\usepackage{graphicx}  
\usepackage{amsmath}
\usepackage{bm}        
\usepackage{amssymb}   
\usepackage{lipsum}
\usepackage{hyperref}
\usepackage{float}
\usepackage{color}

\begin{document}
\title{Role of shape in particle-lipid membrane interactions: from surfing to full engulfment}

\author{Stijn van der Ham}%
\affiliation{Active Soft Matter and Bio-inspired Materials Lab, Faculty of Science and Technology, University of Twente, PO Box 217, 7500 AE Enschede, The Netherlands.
}%

\author{Jaime Agudo-Canalejo}%
\affiliation{ 
Department of Living Matter Physics, \\ Max Planck Insitute for Dynamics and Self-Organization, \\Göttingen, D-37077, Germany
}%

\author{Hanumantha Rao Vutukuri}%
\email{h.r.vutukuri@utwente.nl}
\affiliation{Active Soft Matter and Bio-inspired Materials Lab, Faculty of Science and Technology, University of Twente, PO Box 217, 7500 AE Enschede, The Netherlands.
}%

\begin{abstract}
Understanding and manipulating the interactions between foreign bodies and cell membranes during endo- and phagocytosis is of paramount importance, not only for the fate of living cells but also for numerous biomedical applications. This study aims to elucidate the role of variables such as anisotropic particle shape, curvature, orientation, membrane tension, and adhesive strength in this essential process, using a minimal experimental biomimetic system comprising giant unilamellar vesicles and rod-like particles with different curvatures and aspect ratios. We find that the particle wrapping process is dictated by the balance between the elastic energy penalty and adhesion energy gain, leading to two distinct engulfment pathways, tip-first and side-first, emphasizing the significance of the particle orientation in determining the pathway. Moreover, our experimental results are consistent with theoretical predictions in a state diagram, showcasing how to control the wrapping pathway from surfing to partial to complete wrapping by the interplay between membrane tension and adhesive strength. At moderate particle concentrations, we observed the formation of rod clusters, which exhibited cooperative and sequential wrapping. Our study not only contributes to a comprehensive understanding of the mechanistic intricacies of endocytosis by highlighting how the interplay between the anisotropic particle shape, curvature, orientation, membrane tension, and adhesive strength can influence the engulfment pathway but also provides a foundational base for future research in the field.\\
\\
{\bf{keywords}}:Vesicles, Anisotropic particles, Wrapping,Passive engulfing, Cellular particle uptake
\end{abstract}

\maketitle

Endocytosis is a fundamental cellular process that mediates the uptake of nutrients, pathogens, and therapeutic agents \cite{conner2003regulated}. Understanding and controlling the interaction between foreign bodies and cell membranes in endocytosis is critical for numerous biomedical applications, including targeted drug delivery \cite{singh2009nanoparticle}, intracellular imaging \cite{mitchell2021engineering}, and nanotoxicity studies \cite{gustafson2015nanoparticle}. 
Biomimetic model systems such as giant unilamellar vesicles (GUVs) offer a pathway to a comprehensive understanding of this process \cite{dimova2006practical, dimova2019giant,vutukuri2020active}.
GUVs are extensively used to investigate passive engulfment, a process where particle wrapping is driven by generic physical interactions like particle-membrane adhesion \cite{spanke2020wrapping,ewins2022controlled,eierhoff2014lipid, dinsmore1998hard, dietrich1997adhesion, azadbakht2023wrapping, van2016lipid}.
Particle wrapping is dictated by the balance between the adhesion energy gain from particle-membrane overlap and the elastic energy cost incurred by membrane deformation \cite{dietrich1997adhesion, deserno2004elastic,lipowsky1998vesicles}.

Various studies to date have investigated the role of different parameters influencing the wrapping process, such as particle size \cite{lipowsky1998vesicles,agudo2015critical}, membrane asymmetry \cite{agudo2021particle,agudo2015critical}, membrane tension and fluctuations \cite{deserno2004elastic,ayala2023thermal}, local membrane curvature \cite{agudo2015critical,agudo2015adhesive,AngeloPRL,bahrami2016role}, particle surface properties \cite{ewins2022controlled}, and notably, particle shape \cite{azadbakht2023wrapping, chithrani2006determining, champion2006role, bahrami2013orientational, bahrami2014wrapping, dasgupta2014shape, agudo2020engulfment, huang2013role, zuraw2021membrane, liu2023wrapping, frey2021more}. Particle shape has attracted considerable attention, given its impact on the bending energy cost and potential to modify the engulfment pathway. For example, several theory and simulation studies predict that anisotropic particles, such as ellipsoids and rods, experience a spontaneous rotation during engulfment, a phenomenon that reduces the bending energy cost \cite{bahrami2013orientational, dasgupta2014shape, agudo2020engulfment}. On the other hand, only a limited number of experimental studies have investigated the role of shape anisotropy in the wrapping process, for example, using DNA origami rods \cite{zuraw2021membrane}, microgel particles \cite{liu2023wrapping}, or dumbbell particles \cite{azadbakht2023wrapping} with GUVs, gold nanoparticles with HeLa cells \cite{chithrani2006determining}, and carbon nanotubes with mammalian cells \cite{shi2011cell}. However, these studies are limited by using either non-tunable \cite{liu2023wrapping} and strong binding interactions (e.g., NeutrAvidin-biotin binding \cite{azadbakht2023wrapping}) or relatively small particles \cite{zuraw2021membrane,chithrani2006determining,shi2011cell}, which makes tuning and following their membrane interactions at the single-particle level challenging. A detailed understanding of how rod-like particles interact with membranes during endocytosis is not only of fundamental interest but also has important toxicological implications. This is particularly evident in studies on the size-dependent phagocytosis of asbestos rods by macrophages, which is a critical factor in asbestos-related toxicology research \cite{asbestos,asbestos1}.
Additionally, the role of membrane tension, crucial in processes like inhibiting parasite infection when elevated \cite{cowman2012cellular}, is often neglected in experimental minimal model systems.

In this study, we comprehensively investigate the role of the anisotropic particle shape, curvature, orientation, membrane tension, and adhesive strength in the wrapping pathway at a single-particle level. We employ both straight and curved rods with flat and round tips, along with a non-adsorbing polymer that enables tunable, nonspecific, adhesive interactions via depletion forces \cite{dinsmore1998hard,asakura1954interaction,spanke2020wrapping}. Our model system offers excellent control over the particle's wrapping pathway by the lipid membrane, spanning a range of interactions from a surfing state - where the particle adheres to the membrane without inducing deformation - to partially and fully wrapped states.

We find that the interplay between the elastic energy penalty and the energy gain from depletion attractions controls the wrapping of rod-like particles. Intriguingly, we observe two distinct engulfment pathways: tip-first, where the rod's long axis remains perpendicular to the vesicle membrane, and side-first, where the rod's long axis starts parallel to the membrane and rotates to a perpendicular orientation as the degree of wrapping increases. The initial orientation of the rod relative to the membrane plays a crucial role in determining which pathway the rod follows. 

The paper is organized as follows: we begin by exploring rod-membrane interactions in the high-tension vesicle regime. We then focus on the engulfment pathways of the rods, investigating the effects of rod curvature, aspect ratio, adhesion strength, and membrane tension on these pathways. Following this, we examine the sequential engulfment of clusters involving multiple rods. Finally, corroborating our experimental observations with numerical calculations, we construct a state diagram that encapsulates these findings.

\section{Results and discussion}
Our experimental model system consisted of three main components: giant unilamellar vesicles (GUVs), rod-shaped particles with different curvatures and aspect ratios, and a non-adsorbing polymer. GUVs of 1,2-dioleoyl-sn-glycero-3-phosphocholine (DOPC) lipids were produced with the droplet transfer method \cite{vutukuri2020active,natsume2017preparation}, resulting in vesicles with varying size and membrane tensions.

We employed two fabrication methods to produce the rods from SU-8 photoresist \cite{fernandez2019synthesis,fernandez2020shaping} (see SI Appendix, Section S1 and Fig. S1). The first method produces straight rods with flat tips (Fig. \ref{fig:figure1}a), characterized by an aspect ratio (defined as $h/a$, where $h$ is the length and $a$ is the radius) ranging from 5 to 60. In contrast, the second method generates curved rods with round tips (Fig. \ref{fig:figure1}b). Importantly, the second method can also produce 'straight' rods with round tips, as it allows for the fabrication of a range of curvatures (radius of curvature ranges from approximately $2$ to $50 ~\mathrm{\mu m^{-1}}$ with opening angles from $4$ to $194 ^{\circ}$). Therefore, throughout this paper, we employ 'straight' and 'curved' to refer to the rods' long axis curvature, and we explicitly specify the tip shape when it is necessary for clarity.

In the absence of a non-adsorbing polymer, the rods did not show any specific interactions with the lipid membrane (SI Appendix, Fig. S2). Therefore, to induce an adhesive interaction between the rod and the membrane, we used polyacrylamide (PAM) as a non-adsorbing polymer to serve as a depletant. In the dilute limit, the depletion interaction is given by $E_\mathrm{ad} = n \Delta V k_B T$ \cite{asakura1954interaction}, where $n$ is the number density of depletant and $\Delta V$ is the change in the depletant's excluded volume.
We estimate $\Delta V$ as the overlap volume, $V_\mathrm{ov}$, between the excluded volume regions of the rod and the vesicle membrane. This overlap volume can be calculated using $V_\mathrm{ov} = 2R_G A_\mathrm{co}$, where $R_G$ is the radius of gyration of the depletant and $A_\mathrm{co}$ is the contact area. The adhesive strength, $W$, is given by $W = E_\mathrm{ad}/A_\mathrm{co} = 2R_Gnk_BT$, and can be tuned by varying the PAM concentration (0.25-0.75 wt.-\%). We studied the resulting dynamic interactions between the rods and the membrane using an inverted fluorescence and confocal scanning laser microscope.

\subsection*{Partially wrapped state}
\begin{figure*}[tbhp]
\begin{center}
\includegraphics[width=1.0\linewidth]{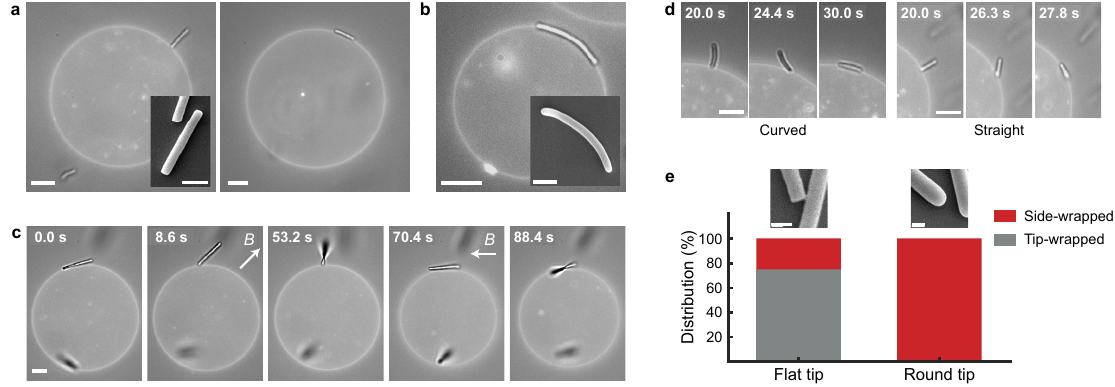}
\end{center}
\vspace{0 cm}
\caption{\textbf{Partially wrapped state.} Combined fluorescence and bright-field microscopy images illustrating the partially wrapped state of rods: a) Flat-tipped rods in the tip-wrapped (left panel) and side-wrapped (right panel) state. b) Round-tipped rod with vesicle-matching curvature in the side-wrapped state. The insets are scanning electron microscope images of a straight and curved rod, respectively. c) Time-lapse images depicting a magnetically responsive flat-tipped rod transitioning from the side-wrapped to the tip-wrapped state and vice versa. The magnetic field is denoted by the letter $B$, and the arrow indicates the direction of the applied magnetic field along which the rod aligns. Between transitions, the magnet was removed to demonstrate the stability of the rod in its new orientation. d) Time-lapse images showing the spontaneous transition of a round-tipped curved and straight rod from the tip-wrapped to the side-wrapped state. e) The distribution of rods over the tip-wrapped and side-wrapped state as a function of tip shape, 5 minutes after their initial adhesion to the vesicle. The scale bars in the microscopy images and insets represent 10 $\mu$m and 1 $\mu$m (a-d), respectively. Scale bars in the SEM image insets (e) represent 0.4 $\mu$m.}
\label{fig:figure1}
\end{figure*}

For high-tension vesicles ($\geq O(10^{-6} ~\mathrm{N/m})$), we observed rods in a partially wrapped state across a range of adhesive strengths (0.25-0.75 wt.-\% PAM). Owing to the high energy penalty for deforming the membrane, these rods only induced a small membrane deformation relative to their unbound state. Interestingly, we observed two distinct configurations in which the rods were partially wrapped by the vesicle membrane, as illustrated in Fig. \ref{fig:figure1}a (Movie S1). In the first configuration, the rod adheres to the membrane with its tip (i.e., short axis), its long axis oriented perpendicular to the vesicle membrane. We refer to this as the tip-wrapped state (Fig. \ref{fig:figure1}a). Conversely, in the second configuration, the rod adheres to the membrane with its side, its long axis parallel to the membrane. We term this as the side-wrapped state (Fig. \ref{fig:figure1}a-b). 

The variation in rod orientation results in a larger membrane-rod overlap area in the side-wrapped state, inducing a stronger adhesive interaction via depletion attractions compared to the tip-wrapped state. As a result, rods in the side-wrapped state bind more strongly to the vesicle membrane than those in the tip-wrapped state. At a low depletion attraction strength ($<$ 0.5 wt.-\% PAM), our observations suggest that rods in the tip-wrapped state are prone to detaching after several seconds due to thermal fluctuations. However, the rods in the side-wrapped state maintain a strong attachment after initial contact. Nonetheless, at higher polymer concentrations ($\geq$ 0.5 wt.-\%), even the rods in the tip-wrapped state maintain a strong attachment to the vesicle membrane. 

We observed spontaneous transitions of rods between the two partially wrapped states occurring only in one direction, from tip-wrapped to side-wrapped. This likely stems from the significantly higher adhesion strength present in the side-wrapped state (the overlap volume in side-wrapped state is 10 times higher than in the tip-wrapped state for $h = 5 \mu$m and $a = 0.2 \mu$m, see SI Appendix, Section S6). Interestingly, the occurrence of this transition and the stability of the tip-wrapped state strongly depended on the shape of the rod's tip. We observed that the majority of rods with round tips spontaneously transitioned from the tip-wrapped to the side-wrapped state due to thermal fluctuations, as depicted in Fig. \ref{fig:figure1}d and Supplementary Movie S2. Conversely, rods with flat tips exhibited greater stability in the tip-wrapped state and infrequently transitioned to the side-wrapped configuration.

To further probe stability, we used magnetically responsive rods, which allowed us to manipulate their position and orientation using an external magnetic field. Using this magnetic field, we positioned the rods in the tip-wrapped state by adhering one of their tips to the vesicle (0.5 wt.-\% PAM). After removing the magnet, we measured the time until a transition to the side-wrapped state occurred. The distribution of rods in the tip-wrapped and side-wrapped states after 5 minutes is depicted in Fig. \ref{fig:figure1}e for rods with round and flat tips, respectively. It should be noted that we monitored the rods' orientation for 10 minutes. However, no further changes in the distribution over partially wrapped states were observed after the initial 5 minutes (see SI Appendix, Fig. S3).

The distribution clearly reveals that rods with round tips transition to the side-wrapped state at a significantly higher rate than those with flat tips. To decouple any effects of lengthwise curvature, we incorporated both straight and curved rods with round tips in our analysis, as illustrated in Fig. \ref{fig:figure1}d and SI Appendix, Fig. S4. The findings indicate no significant dependence on the rods' long axis curvature, suggesting that the tip shape is the decisive factor in determining the stability of the tip-wrapped state.

\begin{figure*}[t]
\begin{center}
\includegraphics[width=1.0\linewidth]{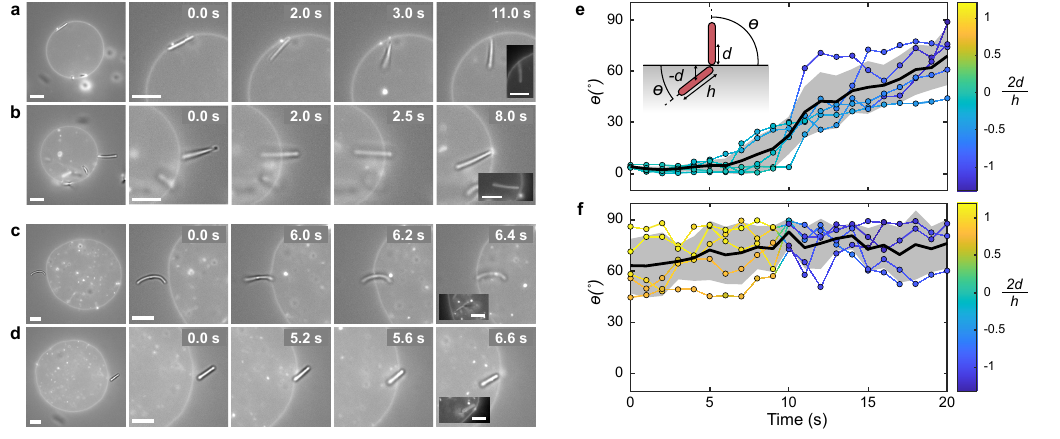}
\end{center}
\vspace{0 cm}
\caption{\textbf{Wrapping pathways.} Time-lapse of combined fluorescence and bright-field microscopy images depicting the engulfment pathways. Insets present the fluorescence image exclusively, showcasing the membrane morphology. a) Side-first engulfment of a round-tipped rod by a GUV: initially in the side-wrapped state, the rod transitions to the fully wrapped state. As the degree of wrapping increases, the rod's long axis rotates from being parallel to becoming perpendicular to the GUV membrane. b) Tip-first engulfment of a flat-tipped rod by a GUV: initially adhered in the tip-wrapped state, the rod transitions to the fully wrapped state. The rod's long axis remains perpendicular to the GUV membrane during this transition. c, d) Tip-first engulfment involving a curved rod with round tips (c), and a straight rod with flat tips (d) by a vesicle under low tension: upon making contact with the membrane, the rod is instantly engulfed, transitioning from the free state to the fully wrapped state. All scale bars represent 5 $\mu$m. e,f) The angle and distance between the rod and the membrane as a function of time during side-first engulfment (e) and tip-first engulfment (f). For each pathway, five independent measurements are plotted. The black line with the grey shaded area represents the average and standard deviation of these measurements, respectively. The inset in (e) provides a schematic depiction of the definitions of the angle $\theta$ and the distance $d$.}
\label{fig:figure2}
\end{figure*}

To delve deeper into how particle shape affects the stability of the partially wrapped state, we again employed magnetically responsive rods, this time manipulating their orientations in order to induce a transition. We found that, with the application of an external magnetic field ($\sim$ 2-10 mT), flat-tipped rods were able to reversibly transition between the side-wrapped and tip-wrapped states, as shown in Fig. \ref{fig:figure1}c (Movie S3). After each transition, the magnetic field was removed to evaluate the stability of the rod in its new configuration. Notably, a rod could transition from the tip-wrapped to the side-wrapped state if it was sufficiently tilted relative to the vesicle membrane. However, transitions from the side-wrapped state to the tip-wrapped state were seldom observed, owing to the high adhesion strength in the side-wrapped state, and necessitated a significant magnetic field strength. 

These experiments indicate the presence of an energy barrier between a stable side-wrapped state and a meta-stable tip-wrapped state, with its magnitude dependent on both the transition direction and the shape of the rod tip. We observe no spontaneous transitions from side-wrapped to tip-wrapped states, implying a substantial energy barrier. However, transitions from tip- to side-wrapped states were frequent for rods with round tips, suggesting a lower energy barrier, comparable to thermal energy. In contrast, flat-tipped rods require rotational manipulation via magnetic force to undergo the same transition, signifying a higher, yet surmountable, energy barrier.

To corroborate our findings, we performed a quantitative analysis to assess the overlap volumes, and thus the adhesive strength, between the rods and vesicle membranes (SI Appendix, Section S6), which consistently supports our qualitative observations. Importantly, this analysis elucidates how the rod's tip shape affects energy barrier heights during transitions from tip- to side-wrapped states; flat-tipped rods, assuming a diagonal orientation with minimal overlap, encountered a higher energy barrier compared to round-tipped rods, which maintained significant overlap throughout the transition.

\subsection*{Fully wrapped state}
We increased the PAM concentration to further probe the role of the adhesive strength in the wrapping process. For vesicles at intermediate tension ($O(10^{-7} ~\mathrm{N/m})$), this increase led to a transition of rods from a partially wrapped to a fully wrapped state at higher adhesive strengths (0.66 – 0.75 wt.-\% PAM). This transition was discontinuous and transpired rapidly within a few seconds or less. Interestingly, we observed that this process occurred along two distinct pathways:

(i) The rod is initially attached in the side-wrapped state and hinges into the GUV (see Fig. \ref{fig:figure2}a and Movie S4). As it transitions from side-wrapped to fully wrapped, the rod reorients itself from parallel to perpendicular relative to the membrane. This reorientation occurs due to the considerable bending energy cost associated with enveloping both highly curved tips. Instead, the rod undergoes a rotational motion during the wrapping process such that only one tip is fully wrapped, while the other remains positioned near the membrane. These findings align with simulation and numerical predictions \cite{bahrami2013orientational, dasgupta2014shape, agudo2020engulfment, huang2013role}, which typically show the engulfment of ellipsoidal and spherocylindrical particles starting with lateral attachment to the membrane and followed by a transition to a perpendicular orientation as wrapping advances. A similar rotational behavior has been observed in experiments involving the wrapping of dumbbell-shaped particles by Azadbakht et al. \cite{azadbakht2023wrapping}. However, it is crucial to distinguish that in their study, rotation is prompted by the inhomogeneous ligand coating distribution on the particle. In contrast, our study clearly demonstrates that the rod's rotation is solely driven by variations in particle curvature, emphasising the significance of curvature in this process.

(ii) Alternatively, the rod begins in the tip-wrapped state, docking to the membrane with its tip, and is then engulfed into the GUV without undergoing reorientation (Fig. \ref{fig:figure2}b and Movie S4). This reflects the findings of Dasgupta et al. \cite{dasgupta2014shape}, who predicted that short rods with flat tips enter tip-first via a rocket-like pathway, whereas rods with higher aspect ratios or more rounded tips exhibit side-first entry, undergoing a rotation throughout the wrapping process. However, our findings extend beyond this model, as we observe also high aspect ratio rods ($h/a$ = 10 - 60) with flat tips stably adhering in the tip-wrapped state and undergoing tip-first entry (Fig. \ref{fig:figure1}a, \ref{fig:figure2}b and SI Appendix, Fig. S6). This aligns with the work of Shi et al. \cite{shi2011cell}, which suggests a similar entry mode for high aspect ratio carbon nanotubes in mammalian cells. We postulate that the increased stability of the tip-wrapped state is due to an energy barrier between the two partial wrapping states, generating a local minimum and enhancing the state's stability.

To quantify the two engulfment pathways, we measured both the angle ($\theta$), and the shortest distance ($d$), between the rod’s centre of mass and the vesicle membrane during engulfment. It should be noted that $\theta$ is considered a positive value regardless of the rod’s orientation relative to the membrane, while $d$ is positive when the rod’s center of mass is outside the vesicle and negative when inside. For more details on the measurement techniques for $\theta$ and $d$, see SI Appendix, Section S2.

A total of five transitions for each pathway were measured, as shown in Fig. \ref{fig:figure2}e-f. The two pathways display distinctive angle progressions: side-first engulfment exhibits a rotation from 0$^{\circ}$ to approximately 45-90$^{\circ}$, while tip-first engulfment maintains rods at an angle between 45-90$^{\circ}$ throughout the process. Our analysis indicates that both pathways are accessible to rods of various lengths and tip shapes. Side-first engulfment is observed for both curved and straight rods, with radii of curvature from 9 to 31 {$\mu$m} and aspect ratios ($h/a$) from 12 to 45. Conversely, tip-first engulfment is predominantly observed for straight rods, with aspect ratios from 10 to 31. Despite the intrinsic error in measuring angles and distances due to the 2D projection of a 3D system, the angle and distance progressions demonstrate remarkable consistency, independent of variations in rod aspect ratio, curvature, and tip shape.

The vesicle membrane morphology when rods are in the fully wrapped state can be inferred from the fluorescence image, as depicted in the inset of Fig. \ref{fig:figure2}a-b. The combined bright-field and fluorescent image distinctly shows both the lipid membrane and the rod, whereas the fluorescent mode exclusively highlights the lipid membrane, owing to the non-fluorescent nature of the rods. This visual evidence affirms that the rod is indeed in a fully wrapped state. Although limitations in spatial resolution make it challenging to definitively identify a membrane neck connecting the rod to the membrane, we observe that the rods remain close to the membrane and are typically oriented at an angle of approximately 45-90$^{\circ}$ to the membrane (Fig. \ref{fig:figure2}e-f). This suggests the presence of a narrow neck connecting them at their tips, inferred from the minimal membrane deformation observed near the rod tip \cite{agudo2016stabilization}.

\subsection*{Cooperative wrapping}
\begin{figure}[thbp]
\begin{center}
\includegraphics[width=1.0\linewidth]{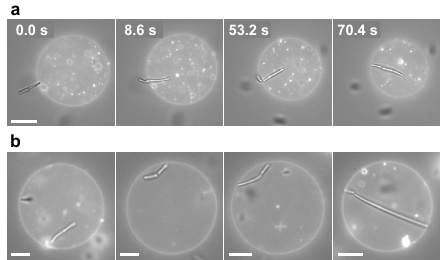}
\end{center}
\vspace{0cm}
\caption{\textbf{Cooperative wrapping of rod clusters.} a) Time-lapse of overlaid fluorescence and bright-field microscopy images revealing the step-wise cooperative wrapping of a two-rod cluster (round-tipped). b) Composite (bright-field + fluorescence) microscopy images of rods oriented tip-to-tip in the fully wrapped state. The scale bars represent 10 $\mu$m.}
\label{fig:figure4}
\end{figure}

At particle concentrations of ($\sim$ 1.0 wt.-\%), we observed rod clusters forming due to non-specific depletion interactions, which promote a side-by-side arrangement resulting from the increased overlap volume between individual rods. At PAM concentrations of 0.5 wt.-\% or higher, these rod clusters exhibited cooperative and sequential wrapping. For instance, a cluster featuring a smaller rod attached to the end of a longer one is depicted in Fig. \ref{fig:figure4}a. Upon contacting a low-tension vesicle, the longer rod was rapidly engulfed, while the shorter rod slid off and remained on the vesicle surface in a side-wrapped state. Intriguingly, after several minutes, the shorter rod transitioned from the side-wrapped to the fully wrapped state, forming a tube-like structure that encapsulated both rods, as shown in Movie S5.

In the final wrapped configuration, the rods are oriented with their highly curved tips pointing towards each other, which minimizes the bending energy penalty as was previously predicted in simulations \cite{raatz2017membrane,xiong2017cooperative}. We note that this configuration, with both rods wrapped within a single membrane tube, is observed frequently. It appears to occur irrespective of the rods' relative aspect ratios, as illustrated in Fig. \ref{fig:figure4}b.

\subsection*{Vesicle membrane tension}
To examine the effect of membrane tension on the wrapping process, we varied the tension of the GUVs. In the low tension regime ($O(10^{-8} ~\mathrm{N/m})$, as measured by shape fluctuation analysis via flickering spectroscopy, see SI Appendix), we observed that full engulfment bypasses the intermediate partially wrapped state. Instead, at intermediate adhesive strengths and higher ($\geq$ 0.5 wt.-\% PAM), rods are spontaneously engulfed by the vesicle upon contact with the membrane. This is demonstrated in Fig. \ref{fig:figure2}c-d and illustrated in Movie S6. Notably, this engulfment behavior, which occurs without the rod needing to reorient, is consistent across both curved and straight rods, regardless of whether an external magnetic field is applied to position the rod near the vesicle membrane.

GUVs with low membrane tension facilitate rod engulfment, while higher tension opposes it. To explore this, we increased the membrane tension via a hypotonic shock (detailed in SI Appendix, Section S3) and monitored rods that were initially in a fully wrapped state at 0.5 wt.-\% PAM. Similar to transitions from the partially to the fully wrapped state, the unwrapping process was sudden and rapid. With increased tension, rods previously in a fully wrapped state transitioned back to a side-wrapped state, necessitating a rotational movement (as shown in SI Appendix, Fig. S7 and Movie S7). We did not observe any transitions to a tip-wrapped state nor complete detachment of the rods from the vesicle, which aligns with expectations given the constant PAM concentration during vesicle inflation. Following the expulsion, the rod rapidly reattached to the vesicle, suggesting that it remained adhered to the membrane.

\subsection*{State diagram}

\begin{figure*}[t]
\begin{center}
\includegraphics[width=1.0\linewidth]{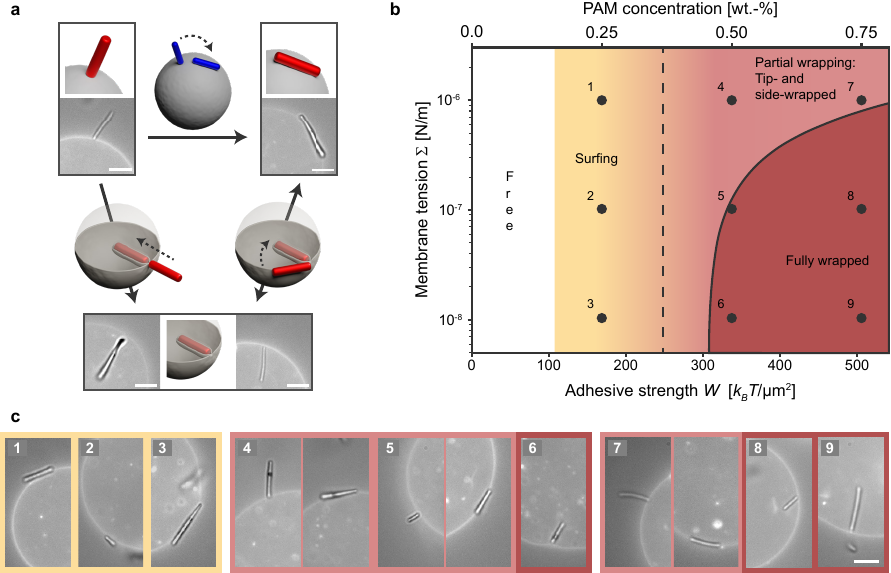}
\end{center}
\vspace{0 cm}
\caption{\textbf{State diagram.}
a) Schematic illustration of transitions between the tip-, side-, and fully wrapped states, with arrows indicating the possible directions of transition pathways. The transition from the tip- to the side-wrapped state was predominantly observed for rods with round tips (depicted in blue), whereas all other transitions were seen for both flat and round-tipped rods (depicted in red).
b) State diagram depicting the experimentally observed and theoretically predicted wrapping states as a function of adhesion strength $W$ and membrane tension $\Sigma$. The vertical dashed line indicates the theoretical transition from the free to the side-wrapped state, as defined by $W>\kappa/(2a^2)$, where $a = 0.2 \ \mu$m is the rod radius. The curved solid line denotes the condition $W>E_{\text{be}}/A + \Sigma$, where $E_{\text{be}}$ is the total bending energy and $A$ represents the area of the rod shape, respectively (SI Appendix, Section S5).
c) Composite (bright-field + fluorescence) microscopy images showcasing the experimentally observed states. Note that these experimental observations qualitatively correspond to the state diagram and are included for illustrative purposes. Their placement within the diagram does not indicate the exact tension of the GUVs; instead, we estimate the tension to be $\Sigma = O(10^{-6} ~\text{N/m}), ~O(10^{-7} ~\text{N/m})$, or $O(10^{-8} ~\text{N/m})$, based on the shape fluctuations of the vesicles (SI Appendix, Fig. S9). For locations 4, 5, and 7, images illustrate both the tip-wrapped state (left) and the side-wrapped state (right). The scale bar is 5 $\mu$m.}
\label{fig:figure3}
\end{figure*}

Figure \ref{fig:figure3} presents a summary of the experimental results for different adhesive strengths and membrane tensions. Figure \ref{fig:figure3}a illustrates the observed pathways between the three states, while Fig. \ref{fig:figure3}b-c displays a state diagram integrating experimental results with numerical calculations.

Following Ref.~\citenum{agudo2020engulfment}, the critical adhesion strength for the transition from the free to the partially wrapped state is given by $W>2\kappa M_\mathrm{pa}^2$, where $W$ is the adhesive strength, $\kappa \approx 20 k_B T$ is the bending rigidity {\cite{vutukuri2020active}}, $M_\mathrm{pa}$ is the local mean curvature of the particle at the point of membrane contact, and we have neglected the local curvature of the membrane, which is much smaller than that of the particle. Since $\kappa$ is a material constant, the critical adhesion solely depends on the local mean curvature $M_\mathrm{pa}$ of the particle. Thus, as we move along the horizontal axis in Fig. \ref{fig:figure3}b from low to high adhesive strength, we expect flat-tipped rods to initially adhere in the tip-wrapped orientation (with $M_\mathrm{pa} \approx 0$, implying $W \gtrsim 0$), followed by round- and flat-tipped rods in the side-wrapped orientation (with $M_\mathrm{pa} = 1/(2a)$ where $a$ represents the rod radius, implying $W > \kappa/(2a^2)$ \cite{weikl2003indirect}) (SI Appendix, Section S5). 

Surprisingly, our experimental observations contradict these predictions, showing that side-wrapping occurs first with increasing adhesion strength (Fig. \ref{fig:figure3}b-c). This discrepancy can be attributed to the interaction range between the rods and the membrane \cite{bickel2003depletion,raatz2014cooperative}. The theoretical prediction above assumes a zero interaction range between the rod and the membrane, while in our experiments, this range is on the order of the depletant size ($R_G \approx 50$ nm). This extended interaction range enables rods to adhere to the membrane without necessitating membrane deformation. Consequently, we observe an intermediate regime between the free and partially wrapped states where the rod adheres to the vesicle membrane in a side-oriented state without deforming the membrane. We will refer to rods in this state as surfing (SI Appendix, Section S6). 

From the partially wrapped state, increasing adhesion strength or decreasing membrane tension leads to a transition to the fully wrapped state. This transition is discontinuous and estimated to occur in the region where $W > E_\mathrm{be}/A + \Sigma$, where $E_\mathrm{be}$ is the bending energy of the membrane wrapped around the rod, $A$ the surface area of the rod, and $\Sigma$ the membrane tension (SI Appendix, Section S5). This transition is essentially an energetic condition and does not account for the possible existence of an energy barrier between the partially and fully wrapped states, which may be quite large and only vanishes at even larger adhesion strengths, as has been shown for ellipsoidal particles in Ref.~\citenum{agudo2020engulfment}. 

Nevertheless, the experimental observations qualitatively align well with these predictions. Notably, within the explored range of depletion strengths, high-tension vesicles ($\geq O(10^{-6} ~\mathrm{N/m})$) do not attain a fully wrapped state, as the membrane tension's energy cost supersedes the gain in adhesion. Conversely, low-tension vesicles ($O(10^{-6} ~\mathrm{N/m})$) achieve a fully wrapped state at intermediate adhesion strengths and above ($\geq$ 0.5 wt.-\% PAM). Finally, for vesicles with intermediate tension, achieving a fully wrapped state necessitates higher adhesion strengths (above $0.5$ wt.-\%).

It is worth noting that the simulation and theoretical works often assume that the GUV membranes have zero spontaneous curvature. However, in actual experimental conditions, this is not always the case, as inferred by the observation of inward tubular structures in some of our vesicles. Nevertheless, large spontaneous curvatures (with magnitudes much larger than the inverse particle size) can be approximated as generating an effective ``spontaneous tension'' \cite{agudo2021particle}. Indeed, we find that our experimental results are qualitatively consistent with the theoretical predictions for nonzero tension but zero spontaneous curvature.

\section*{Conclusion}
In conclusion, our research elucidates the critical role of anisotropic particle shape, orientation, curvature, membrane tension, and adhesive strength in the engulfment process. We demonstrate how to control the particle engulfment, spanning a range of interactions from surfing to partially and fully wrapped states, orchestrated by the interplay between non-specific adhesive strength and the elastic energy penalty. 
In the partially wrapped state, an intriguing finding is the role of tip-shape of the rod in determining the stability of the tip-wrapped state, highlighting the importance of the curvature of the tip in the wrapping process. Our quantitative analysis reveals that rods with flat tips exhibit stable adhesion in this state, while those with rounded tips tend to transition to the side-wrapped state. 

Our study identifies two distinct engulfment pathways by which rod-shaped particles can achieve a fully wrapped state: rods that initially adhere with their tips follow a tip-first path, while those adhering laterally take a side-first path, involving a rotation of the rod as the degree of wrapping advances. Notably, our findings indicate that the angle progression of rods during engulfment remains consistent over a range of rod aspect ratios and curvatures, indicating that the pathways are determined by particle shape anisotropy rather than the rod's long-axis curvature and aspect ratio. Remarkably, under very low membrane tension conditions, rods directly pursue a tip-first pathway to a completely wrapped state, irrespective of their aspect ratio and curvature, emphasizing the crucial influence of membrane tension in the engulfment process. Furthermore, we find that when multiple rods are in a fully wrapped state, they reach a tube-like structure, pointing their highly curved tips toward each other to minimize the bending energy. Our experimental results are consistent with theoretical predictions in a state diagram, illustrating how to control the wrapping pathway from surfing to partial to complete wrapping by modulating membrane tension and adhesive strength. To the best of our knowledge, this study is novel in its precise manipulation of the engulfment route of anisotropic rod-like particles with varying aspect ratios and curvatures.

Overall, our research contributes to a broader understanding of anisotropic particle engulfment pathways and their significance in endocytosis and phagocytosis \cite{safari2020neutrophils}, extending implications for advanced biomedical applications such as targeted drug delivery, intracellular imaging, and the critical area of nanotoxicity studies \cite{asbestos,asbestos1}. While our experiments employed a simplified biomimetic system, the principles we have established pave the way for future research in more complex environments, potentially incorporating diverse lipid compositions, membrane proteins, and intricate particle geometries to further dissect the complexities of particle engulfment.

\bigskip

\section{Data availability}
All data needed to evaluate the conclusions in the paper are present in the paper and/or the Supplementary Materials. Additional data related to this paper are available from the corresponding author upon reasonable request.


\section{Acknowledgements}
We thank Ineke Punt for SEM measurements, Frieder Mugele and Mireille Claessens for kindly providing access to confocal and fluorescence microscopes. H.R.V. acknowledges funding from the Netherlands Organization for Scientific Research (NWO, Dutch Science Foundation).

\section{Author contributions}
H.R.V. conceived the project. H.R.V. and S.v.d.H. designed the research. S.v.d.H. performed all the experimental work and analysis. H.R.V. supervised the research. J.A.-C. performed the numerical calculations. All authors participated in writing, the discussions, and reviewed and edited the manuscript.

\section{Competing Interests}
The authors declare no competing interest.

\bibliographystyle{apsrev4-1}
\bibliography{Bibliography}
\end{document}